\documentclass[aps,prd,preprintnumbers,showpacs,showkeys,nofootinbib,
superscriptaddress,fleqn,floatfix,tightenlines,10pt,
byrevtex]{revtex4-2}
\usepackage{amsmath,amsfonts,amssymb,amscd,amsxtra,amsthm}

\usepackage{appendix}
\usepackage[unicode]{hyperref}
\usepackage{graphicx,subfigure}
\graphicspath{{./Figures/}}
\usepackage{indentfirst}
\usepackage{float}
\usepackage{bm}
\usepackage{physics}
\usepackage{multirow}
\usepackage{makecell}
\usepackage{ltablex}
\usepackage{siunitx}

\usepackage[utf8]{inputenc}
\usepackage[normalem]{ulem} 
\graphicspath{{./Figures/}}
\usepackage{indentfirst}
\usepackage{float}
\usepackage[dvipsnames]{xcolor} 

\renewcommand{\det}{\mathrm{det}}

\newcommand{\bea}{\begin{eqnarray}}
\newcommand{\eea}{\end{eqnarray}}

\newcommand{\re}[1]{(\ref{#1})}
\usepackage[compat=1.1.0]{tikz-feynman}
\usepackage{contour}
\usetikzlibrary{shapes,arrows,positioning,automata,backgrounds,calc,er,patterns}
\tikzfeynmanset{every blob={/tikz/fill=none, /tikz/inner sep=1pt}}
\makeatletter

\begin{document}

\title{Dark energy and QCD instanton vacuum\\
   in Friedmann–Lemaître–Robertson–Walker  universe}

\author{M.~M. Musakhanov}
\email{musakhanov@gmail.com}
\affiliation{Institute of Theoretical Physics, National University of Uzbekistan,
Tashkent 100174, Uzbekistan}

\begin{abstract}

The standard model of the universe, $\lambda$CDM, is based on the Friedmann-Lemaître-Robertson-Walker metric with a flat three-dimensional coordinate space and the Friedmann equations~\cite{ParticleDataGroup:2024cfk}.
 {The cosmological constant $\lambda$ provides the cancellation of the matter field contributions in the flat (Minkowski) space, as 
was proposed long ago in 1967 by Zeldovich  for the
first time to our knowledge~\cite{Zeldovich:1967gd,Zeldovich:1968ehl}.
The dynamical dark energy appears on the surface of the vacuum energy of matter fields
at the flat (Minkowski) space.}
 Within the Standard Model, the gluon Yang-Mills (YM) fields are playing a specific role since the properties of their vacuum, where there is 
 the presence of the gluon condensate, provide the nonperturbative vacuum energy. It is natural to apply the successful instanton
 liquid model (ILM) of the QCD vacuum and its lowest excitations.   Our aim is to calculate the contribution of gluon YM fields to the dark energy density. 
We find that the universe metric is generating the QCD vacuum excitation, which 
gives the contribution to the dark energy density. But this one 
may hardly play a central role in the dynamics of the universe, since its timescale is too small. 
We also find the equation-of-state parameters
$w_0=-1,\,\,\,w_a=0$ in accordance with $\lambda$CDM, while the newest data, 
analyzed at~\cite{Shajib:2025tpd},  
give them at least in the range
$-0.91 <w_0< -0.73,\,\,\,\,-1.05< w_a <- 0.65$. They are   requesting a
contribution from an
ultralight scalars such an axions, 
 { or from 
YM field topological configurations 
with the nontrivial holonomy due to the deviation from a 
pure de Sitter state ~\cite{VanWaerbeke:2025shm}.
}

\end{abstract}

\date{\today}
\maketitle

\section{Introduction}\label{intro}
 The impact of matter (particles and fields) on cosmological observables of the universe can be quantified in terms of its equation of state $w$, the ratio of its pressure $p$ to its energy density $\epsilon$, $w=p/\epsilon$. A phenomenological description of the redshift (time) evolution is represented by a two-parameter 
function $w(z)=w_0+w_a\,z/(1+z)$ where $w_0 \sim$ the present value, 
$w_a\sim$ the time evolution, and $z$ is the redshift. 
The standard cosmological $\lambda$ cold dark matter ($\lambda$CDM) model, with $70\%$ vacuum energy (equivalently the cosmological constant $\lambda$) and $30\%$ matter (CDM plus baryons),  is fixing $ w_0=-1$ and $w_a=0$.

$\lambda$CDM is based on the Friedmann–Lemaître–Robertson–Walker  (FLRW) space metric:
\bea\label{FLRWmetric}
&&ds^2=dt^2-a^2(t)d\vec x^2=a^2(\tau)(d\tau^2-d\vec x^2),\,\,\, dt=a(\tau)d\tau,
\\\nonumber
&&g_{00}=g^{00}=1, g_{mn}=-a^2(t)\delta_{mn},\,\,\,g^{mn}=-a^{-2}(t)\delta_{mn},
\,\,\, g\equiv \det g_{\mu\nu}=-a^6(t).
\eea
$dt$ is proper while $d\tau=dt/a(t)$ is conformal time. 
We define the FLRW space with conformal time $\tau$ as FLRW conformal space.
The basic Friedmann equations of the FLRW universe 
(see recent PDG review~\cite{ParticleDataGroup:2024cfk} ) are 
\bea
H^2=\left(\frac{\dot a}{a}\right)^2 = \frac{8\pi G}{3} \epsilon_{tot}+\frac{\lambda}{3},
\,\,\,\,
\frac{\ddot a}{a} = -\frac{4\pi G}{3}(\epsilon_{tot}+3p_{tot})+\frac{\lambda}{3},
\label{Friedmann}\eea
where $\epsilon_{tot}$ is the total energy density of the universe and $p_{tot}$ 
is the total pressure. 
 Here and at the following, we are using $\hbar=c=1$, 
 Planck mass $M_{\rm Pl} = 1.2 \, 10^{22}$~MeV, 
 Newton's gravity constant $G = M_{\rm Pl}^{-2}$, present time 
Hubble parameter $H_0={\dot a(t_0)}/{a(t_0)}
\approx 1.4\times 10^{-39}
 $~MeV~\cite{ParticleDataGroup:2024cfk}. Also, we will use the definition of metric field $h(x)$ as $a(x)=\exp(h(x)/2)$.

 { The total energy density and pressure of the Universe has an essential part given by the Standard Model (SM) matter fields vacuum contributions as perturbative vacuum fields fluctuations and nonperturbative
Higgs potential and QCD vacuum condensates. Current data indicate that all 
vacuum  contributions are the largest contributors to the cosmological energy density budget ($0.685 \pm 0.007$)~\cite{ParticleDataGroup:2024cfk} 
and we for a moment will neglect by the contributions of the 
pressureless matter and relativistic particles.
  
In the flat (Minkowski) space, the left sides of Eqs.~\re{Friedmann} 
are zero, which correspond to the total compensation of the flat (Minkowski) space matter fields vacuum contributions  $\epsilon_{\rm vac}(0)$ by the  cosmological constant $\lambda$  and $\epsilon_{\rm vac}(0)+p_{\rm vac}(0)=0$. 
 It means at least that gravitating  dynamical dark energy 
is appearing on top of the vacuum energy of matter fields 
at the flat (Minkowski) space. Such an interpretation of the  cosmological constant $\lambda$ was proposed long ago in 1967 by Zeldovich  for the
first time to our knowledge~\cite{Zeldovich:1967gd,Zeldovich:1968ehl} and are  widely
accepted now~\cite{Bjorken:2001pe,Bjorken:2004an,Schutzhold:2002sg,Klinkhamer:2007pe,Klinkhamer:2008ns,Urban:2009vy,Urban:2009yg,Zhitnitsky:2013pna,Zhitnitsky:2015dia,Barvinsky:2017lfl,VanWaerbeke:2025shm}.

}

Certainly, first of all, we have to take into account Standard Model fields.
Among them, the gluon Yang-Mills (YM) fields are playing 
 { the specific role since the properties of their vacuum, where the presence of the gluon condensate and YM conformal (scale) anomaly, 
provide the nonperturbative vacuum energy.}
 
In QCD vacuum perturbative quantum fluctuations 
are living over spontaneously created nonperturbative classical YM fields, which are the paths of the quantum 
tunneling processes -- (anti)instantons, randomly populating the total space. These fields are giving nonperturbative contribution $\epsilon_{\rm vac}$  to the vacuum energy density of YM fields.
The nonperturbative part of the  energy-momentum tensor vacuum expectation 
in the space with the metric $g_{\mu\nu}$  is given by
$<T_{\mu\nu}>_{\rm np}=\epsilon_{\rm vac} g_{\mu\nu}.$
The vacuum energy density $\epsilon_{\rm vac}$  can be taken from the trace of this tensor
as $<T_\mu^\mu>_{\rm np}=4\epsilon_{\rm vac}$.  
This trace in QCD is nonzero only due to
the quantum loops, which are generating scale anomaly. This one is related 
to the renormalization of quantum loops, which leads to the  the 
dimensional transmutation  and 
provides dimension-full parameter $\Lambda_{\rm QCD}$.

 In the flat (Minkowski) space, the fields vacuum energy density 
 $\epsilon_{\rm vac}(0)$
is canceled by the cosmological constant 
$\lambda$ on the right side of the Friedmann equations~\re{Friedmann}. 
In the FLRW space it was assumed~\cite{Schutzhold:2002sg} 
the metric depended on an additional  contribution
$\sim O(H\Lambda^3_{\rm QCD})$ induced by gluon condensate. But this assumption 
 was ruled out at the work~\cite{Shapiro:2008sf}. 

 {More detailed further studies  of the QCD in a 
curved space~\cite{Urban:2009vy,Urban:2009yg,Zhitnitsky:2013pna,Zhitnitsky:2015dia,Barvinsky:2017lfl,VanWaerbeke:2025shm} showed that such type
 contribution can appear due to the 
YM topological configurations with the nontrivial holonomy, 
compactified to a Euclidean time circle $\mathbb{S}^1_{\kappa^{-1}}$
with the radius $\kappa^{-1}$. 
This nontrivial holonomy is a source of
$O(\kappa\Lambda^3_{\rm QCD})$ contribution to the YM vacuum energy.
This phenomena is quit similar nonzero temperature QCD 
(KvBLL calorons)~\cite{Kraan:1998pm,Lee:1998bb}. 
It is important that these deviations
from a pure de Sitter state can explain the 
current data from the DESI Collaboration~\cite{DESI:2025zgx} as well as other data.}
The analysis of them~\cite{Shajib:2025tpd},  
give the equation-of-state parameters in the range
$-0.91 <w_0< -0.73$ and $-1.05< w_a <- 0.65,$ which
is also a well corresponding phenomenological
 massive scalar field model with the mass just
slightly below the current value of the Hubble parameter, 
for instance, $m\approx 0.82 H(t_0)=1.2\times 10^{-39}$~MeV
for a free, massive scalar), corresponding to a current value
of the equation-of-state parameter $w_0\approx-0.9$~\cite{Shajib:2025tpd}).
Such an ultralight scalar mass is a natural in a model like QCD axions.

The name dark energy has become usual for such types 
of contributions.

In the present work we are staying within the $\lambda$CDM 
FLRW metric~\re{FLRWmetric}.  
In this case, it is natural to apply  
the successful instanton liquid model (ILM) of the QCD vacuum and its lowest excitations.  
Originally, ILM was formulated in Euclidean flat space, taking into account essential topological 
properties of the YM fields~\cite{Shuryak:1981ff,Diakonov:1983hh} (see reviews~\cite{Diakonov:2002fq,Shuryak:2018fjr}). 
We are considering only gluon YM fields as a matter field living 
at nonflat space with the FLRW metric~\re{FLRWmetric}. 
Our framework is  the ILM extended to this space, which is described in Sec.~\ref{YM}. It is important here that perturbative loops are generating
the dimensional transmutation and essential for us 
the metric - gluons interaction term, leading to the  
 conformal (scale) anomaly properties of QCD. These properties
 are presented by scalar glueball field effective lagrangian,
which is describing the fluctuations of instantons density.
In Sec.~\ref{DE} the effective lagrangian is used for the
calculations of the energy density and the pressure of the excitations of the 
scalar glueball field $\xi$ induced by the metric $h$, which is  
YM dark energy density $\epsilon_{\rm dark}(h)$ and pressure $p_{\rm dark}(h) $ 
of the FLRW universe. 

\section{Gluon YM fields in the FLRW universe }
\label{YM}

We take the YM action in  Weyl gauge ( $A^a_0 = 0$) at FLRW space~\re{FLRWmetric}, 
 change the time $t$ to the conformal time, $\tau$ and have 
\bea\label{YMaction}
&&S=-\frac{1}{4e^2}\int d^4x (-g)^{1/2} F^a_{\mu\nu}F^{a\mu\nu}
=\frac{1}{2e^2}\int d\tau\int d^3x  ( {\vec E}^{a2}-{\vec B}^{a2}), 
\\\nonumber
&& E_i^a=\frac{\partial A_i^a}{\partial\tau},\,\,\, 
B_i^a=\frac{1}{2}\epsilon_{ijk}(\partial_j A_k^a-\partial_k A_j^a+\epsilon_{abc}A_j^b A_k^c )
\eea
We see the YM action in the FLRW conformal space $(\tau,\vec x)$ completely coincide with the action in the flat (Minkowski) space $(t,\vec x)$. We expect that YM fields in the  FLRW conformal space have the same properties as in the 
flat (Minkowski) space.

\subsection*{Conformal time}
The conformal time $\tau$ in FLRW space, filled by constant 
positive density energy, 
at present time $t_0$ is defined by the solution of the Friedmann equation~\re{Friedmann} 
$a(t_0)\sim\exp(Ht_0)$ and for the whole period can be taken as  
$a(t)= \sinh (Ht)$.
Then, the conformal time is given by
\bea
\tau(t)=\int^t_{t_0}\frac{dt'}{a(t')}
=\frac{1}{ H}\int^{Ht}_{Ht_0} \frac{dx}{\sinh (x)}
=-\frac{1}{2H}\ln\left(\frac{e^{Ht'}+e^{-Ht'}+2}{e^{Ht'}
+e^{-Ht'}-2}\right)_{t'=t_0}^{t'=t} .
\label{tau}\eea 
We see 
$\tau(t\rightarrow 0)\rightarrow -\infty,\,\,\,  
\tau(t\rightarrow \pm\infty)\rightarrow 0 $.
So, $\tau=-\infty$ correspond to the initial time $t=0$, while $\tau=0$ -- final one $t=\infty.$

We will also need the conformal time in Euclidean FLRW space. 
Define $t=i\tilde t$, $\tilde x=(\tilde t,\vec x)$ and the metrics $\tilde g_{\alpha\beta} $ with Euclidean signature $(++++)$ 
as $g_{\mu\nu}( x)=-\tilde g_{\alpha\beta}(\tilde x)\frac{\partial \tilde x_\mu}{\partial x_\alpha}\frac{\partial \tilde x_\nu}{\partial x_\beta}$. Then$-\det g_{\mu\nu}\equiv -g=\tilde g$.

The FLRW Euclidean metric is given by 
\bea
d\tilde s^2=d\tilde t^2+ a^2(\tilde t)d\vec x^2= a^2(\tilde\tau)(d\tilde\tau^2+d\vec x^2),
\label{FRWE}\eea
where $d\tilde t$ is proper while $d\tilde\tau=d\tilde t/a(\tilde t)$ is conformal times. At $a=1$ we have Euclidean flat metric.
The conformal Euclidean time $\tilde\tau$ is given by
\bea
\tilde\tau(\tilde t)=\int^{H\tilde t}_{\pi/2}\frac{d\tilde t'}{\sin(H\tilde t)}
=-\frac{1}{2 H}\ln\left(\frac{1+ \cos H\tilde t}{1- \cos H\tilde t}\right).
\label{tildetau}\eea
We have
 $\tilde\tau(H\tilde t\rightarrow 0)=-\infty,
\,\,\,\,
 \tilde\tau(H\tilde t\rightarrow \pi)=+\infty  .
$

\subsection*{Topology of YM fields in FLRW space}

We expect that the topology of YM fields will be the same in both spaces
--flat and FLRW conformal ones.
We  introduce in FLRW conformal space the topological collective coordinate 
of the YM fields--the Chern-Simons (CS) number as
 \bea
 N_{\rm CS}(\tau)=\frac{1}{16\pi}\int d^3x \epsilon^{ijk}\left(A_i^a\partial_j A_k^a   
 +\frac{1}{3}\epsilon_{abc} A_i^a A_j^b A_k^c\right)
\label{NCS} \eea
and take a large  gauge transformation of the YM fields
$A_i\rightarrow U^+(A_i+i\partial_i)U$
 with topological  winding number 
\bea
N_W=\frac{1}{24\pi^2}\int d^3x\epsilon^{ijk}\left[ (U^+\partial_i U) 
(U^+\partial_j U) (U^+\partial_k U)   \right].
\label{NW}\eea
 with $SU(2)$ matrix $U$, which is providing mapping of three-dimensional $SU(2)$ internal space $S^3$ to the external coordinate space $S^3$. Under this transformation
 the CS number is changed as
 \bea
 N_{\rm CS}\rightarrow  N_{\rm CS} + N_{W}.
\label{NCS1} \eea
Let us take the potential energy of YM fields as a function of CS number  $N_{\rm CS}$
\bea
V(N_{\rm CS})=\frac{1}{2e^2}\int d^3x ({\vec B}^{a}(\tau,\vec x))^2 .
\label{V}\eea 
  We arrive at the conclusion first pointed out in~\cite{Faddeev:1976qt,Faddeev:1976pg,Jackiw:1976pf}, to the best of knowledge, 
  that the potential $V(N_{\rm CS})$ is periodic on the collective coordinate   $N_{\rm CS}$ like a one-dimensional crystal
and oscillatorlike in all other directions in functional space.
The quasiclassical quantization leads to the band structure of the energies. 
The width of the band is defined by the amplitude  of tunneling
$\sim e^{-S_E}$, where the action $S_E$ is calculated on the Euclidean classical trajectory between the nearest minima.
 This trajectory is a YM (anti)instanton with topological charge $Q_T=\pm 1$,
  which is the solution of YM equations in Euclidean FLRW conformal space. 

\subsection*{(Anti)instanton in conformal Euclidean  FLRW space }

 Euclidean YM action in
Euclidean  conformal FLRW metric 
$\tilde g_{\mu\nu}= \exp( h(\tilde\tau))\delta_{\mu\nu}$   and at Weyl gauge $A_0=0$ can be easily found from~\re{YMaction} as
\bea\label{YMactionFRWE}
S_E
=\frac{1}{2e^2}\int d\tilde\tau\int d^3x  ( {\vec E}^{a2}+{\vec B}^{a2}), \,\,\, 
 E_i^a=\frac{\partial A_i^a}{\partial\tilde\tau},\,\,\, 
B_i^a=\frac{1}{2}\epsilon_{ijk}(\partial_j A_k^a-\partial_k A_j^a+\epsilon_{abc}A_j^b A_k^c ),
\eea
which completely coincides with the YM action in Euclidean flat space. 
It means we may take the known solution~\cite{Dunne:2000if} for the
 (anti)instanton $A^\pm$ with $Q_T=\pm 1$ in singular gauge.  
In the general case, the solution is given by 
$U_i^+ A^\pm_{i}(\tilde\tau-\tilde\tau_i,\vec x-\vec x_i)U_i $ 
the collective coordinates:  four-dimensional position $(\tilde\tau_i,\vec x_i)$, 
  color orientation, $U_i$ and  size $\rho_i$.

\subsection*{Instanton liquid model of QCD vacuum in FLRW conformal space }

ILM in the flat space for the QCD vacuum and its lowest excitations 
-- hadrons~\cite{Shuryak:1981ff,Diakonov:1983hh} (see reviews~\cite{Diakonov:2002fq,Shuryak:2018fjr} ) can easily translated to the  FLRW conformal space. 
The quantum tunneling processes, randomly distributed over  FLRW conformal space, are corresponding to the sum ansatz for the total YM field 
\bea
A^{\rm ILM}_i(\tilde\tau,\vec x) =\sum_\pm\sum_{i=1}^{N_\pm} U_i^+ A^\pm_{i}(\tilde\tau-\tilde\tau_i,\vec x-\vec x_i)U_i.
\label{AILM}\eea
The variational~\cite{Diakonov:1983hh} and lattice~\cite{Chu:1994vi} 
calculations and phenomenological
estimations~\cite{Shuryak:1981ff} demonstrated the concentration 
of the (anti)instanton sizes  around an average of one $\rho$
 as an effect of instanton-(anti)instanton interactions. 
In the following, we take  the same size $\rho$ 
for all (anti)instantons. Also, we take $N_+=N_-=N/2$, since $CP$ invariance is requesting this one. Another important parameter of ILM is the interinstanton distance $R$, which is certainly related to the density of instanton media as $N/V=1/R^4$.  The renormalization properties of the QCD are controlling the dispersion of the
number of instantons $N$. At color number $N_c\rightarrow\infty$, 
the dispersion becomes zero~\cite{Diakonov:1995qy}.  
The various estimations of ILM main parameters  
(variational~\cite{Diakonov:1983hh}, phenomenological~\cite{Shuryak:1981ff} and
lattice~\cite{Chu:1994vi}) give $\rho\approx 0.33$~fm and $R\approx 1$~fm.

Finally, ILM's remaining problems are the integrations over 
the (anti)instantons collective coordinates [four-dimensional position $(\tau_i,\vec x_i)$ and color orientation $U_i$] and over quantum fluctuations living above the classical background~\re{AILM}. 
The calculation of these perturbative quantum 
corrections--gluon (and ghost) loops--is requesting the renormalization 
of them at the specific scale, which is in ILM given by 
the instanton size $\rho$. 
As a result, we have the dimensional transmutation, which provides dimension-full QCD parameters    $\Lambda_{\rm QCD}$ as
\bea
\frac{2\pi}{\alpha_s(\rho)}=b_1\ln\frac{1}{\Lambda_{\rm QCD}\rho},\,\,\,\,
\alpha_s(\rho)=\frac{e^2(\rho)}{4\pi}.
\label{lambdaqcd}\eea 
Here one-loops provide $b_1=11N_c/3$ (in QCD $N_c=3$) 
and $\Lambda_{\rm QCD}\approx 100 $~MeV.
This dimensional transmutation leads to the
QCD conformal (scale) anomaly, and we have 
non-zero trace of the 
YM energy-momentum tensor, which is in the ILM given by
\bea
<T^\mu_\mu>_{\rm ILM}\approx -b_1\frac{<F^{a\mu\nu}F^a_{\mu\nu}>_{\rm ILM}}{32\pi^2}
= -b_1/R^4
\label{traceQCD}\eea
It is obvious that
the nonperturbative QCD vacuum energy in the flat (Minkowski) space case 
\bea
\epsilon_{\rm vac}(0) =<T^\mu_\mu>_{\rm ILM}/4 = -\frac{11}{4} \frac{1}{R^4}
\approx-550~{\rm MeV/fm^3}
\label{ILMvacuumenergy}\eea
From previous discussion, we remember that $\lambda$ must completely cancel 
$\epsilon_{\rm vac}(0)$ in the right sides of the Friedmann equations. 
It looks strange that the FLRW metric did not modify the YM fields energy density.

\subsection*{ YM effective action in $D=(4-\varepsilon$)-dim  Euclidean FLRW conformal space }
To clarify the above-mentioned problem, let us consider YM fields in
the  Euclidean  FLRW conformal D-dimensional space with the  metric [where $\tilde x=(\tilde\tau,\vec x)$]
$
\tilde g_{\mu\nu}(\tilde x)=\exp(h(\tilde x))\delta_{\mu\nu}.
$
The classical YM action has $h$ dependence as 
\bea
S=\frac{1}{4e_0^2}\int d^D\tilde x(e^{h/2})^{D-4} F^a_{\mu\lambda}F^a_{\mu\lambda}
\label{SD}\eea 
and dimensional regularization $\epsilon=4-D$ of the one-loop 
contributions~\cite{Migdal:1982jp} leads to the renormalization and 
dimensional transmutation~\re{lambdaqcd} as
\bea
\frac{1}{e_0^2}=\frac{1}{e(\rho)^2}\left(1+\frac{b_1}{8\pi^2}\frac{e(\rho)^2}{\epsilon}\right)
\label{regularization}\eea
There, the second term represents  gluon (and ghost) loops contribution. So, we have the
effective action $S_{\rm eff}$ as
\bea
&&S_{\rm eff}=\frac{1}{4e^2(\rho)}\int d^D\tilde x(e^{h/2})^{D-4} F^a_{\mu\lambda}F^a_{\mu\lambda}+\Delta S, 
\nonumber\\
&&\Delta S=\lim_{\varepsilon\rightarrow 0}\int d^4\tilde x 
[(e^{h/2})^{-\varepsilon}-1]\frac{1}{\varepsilon}
\frac{b_1 }{32\pi^2}F^a_{\mu\lambda}F^a_{\mu\lambda}
=-\int d^4\tilde x\frac{h}{2}\frac{b_1}{32\pi^2} F^a_{\mu\lambda}F^a_{\mu\lambda},
\label{deltaS}\eea
where the gluon fields are interacting with the external field $h$ in the
 conformal FLRW space. Their coupling is given by $\Delta S$. 
It is natural to assume that a metric field $h$ can generate 
a scalar glueball field as the 
excitations of instanton liquid with vacuum quantum numbers. 

\subsection*{Scalar glueball field as excitations of instanton liquid in the conformal  FLRW space}

Fluctuations of the numbers of instantons in the flat space are controlled by the renormalization (scale dependence) properties of 
QCD~\cite{Novikov:1981xi,Diakonov:1995qy}.

Again we explore one-to-one correspondence of the flat space effective action to the 
same one in the  conformal  FLRW space. 
The partition function $Z$ in that space with the account of the
Eqs.~(\ref{SD},\ref{deltaS}) is given by
\bea
Z=\int DA \exp\left(-S_{\rm eff}\right)
=\int DA \exp\left(-\frac{1}{4e^2(\rho)}\int d^4\tilde x F^a_{\mu\lambda}F^a_{\mu\lambda}-\Delta S\right),
\label{Z}\eea
Since in QCD we have only one dimensional-full parameter, $\Lambda_{\rm QCD}$
the vacuum average of $F^2= F^a_{\mu\lambda}F^a_{\mu\lambda}$ which is given by 
\bea
< F^2>=-\frac{1}{V}
\frac{d\ln Z}{d(\frac{1}{4e^2(\rho)})}
\sim \Lambda_{\rm QCD}^4
=\rho^{-4}\exp\left(-\frac{1}{b_1}
\frac{32\pi^2}{e^2(\rho)}\right)
\label{theorem}\eea
is defined only by its dimensionality.
Repeating the differentiations of both sides of~\re{theorem} we get the (infinite) series of 
low-energy theorems for the series of  $F^2$ correlators. 
These simple arguments can even be extended to the similar series of
the correlators of local colorless  QCD operators with $F^2$ 
operators~\cite{Novikov:1981xi} (see review~\cite{Shifman:1988zk}).

As a result, the fluctuations of the instanton density $N/V$ are controlled by these theorems~\cite{Diakonov:1995qy}. 
These features of the QCD can be reproduced by means of corresponding 
 scalar glueball field effective Lagrangian
  $L(\xi)$~\cite{Migdal:1982jp,Ellis:1984jv,Kharzeev:2004ct}, 
 where scalar glueball field  is defined as 
 \bea
 |\sigma_0|\exp\xi=-T^\mu_\mu=\frac{b_1}{32\pi^2} F^2
\label{xi} \eea 
This field $\xi$ corresponds to the
density fluctuations of the ILM instanton liquid.
 The kinetic part $L(\xi)$ must be scale invariant, 
 while the potential part must be 
 scale  noninvariant in to reproduce QCD scale 
 anomaly and mentioned 
 low-energy theorems~\cite{Shifman:1988zk,Migdal:1982jp}. 
 In the ILM  $|\sigma_0|=-4\epsilon_{\rm vac}(0)=b_1/R^4$.
 So, we have a Lagrangian $L(\xi)$ for the real scalar glueball field $\xi$  
 with glueball mass $m$, which realize low-energy theorems, as
\bea
L(\xi)=\frac{|\sigma_0|}{4m^2} \frac{1}{2} (\partial_\mu\xi)^2\exp{\frac{1}{2}\xi}+V(\xi),\,\,\,
V(\xi)=\frac{|\sigma_0|}{4}(\xi-1)\exp\xi
\label{Lxi}\eea 
\centerline{\includegraphics[scale=0.5]{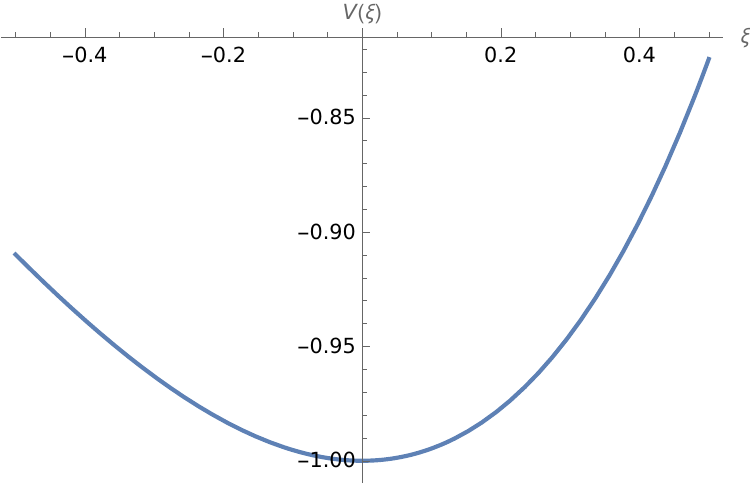}} 

\section{Glueball dark energy density and pressure of the FLRW universe}
\label{DE}

Let us consider the effective action $S_{\rm eff}[\xi,h]$ in the conformal 
FLRW space with Euclidean time $\tilde\tau$.
The effective action has  the term $\Delta S$ describing the interaction of the metric field $h$ with the scalar glueball field $\xi$.
With the account of the $\Delta S$, Eq.~\re{deltaS}, the effective action 
 $S_{\rm eff}[\xi,h]$ and the partition function $Z$ become 
\bea\nonumber
&&S_{\rm eff}[\xi,h]=\int d^3x d\tilde\tau \left(L(\xi)
-\frac{h}{2}|\sigma_0|\exp\xi\right)
=\int d^3x d\tilde t \exp(-h/2) \left(L(\xi)
-\frac{h}{2} |\sigma_0|\exp\xi\right),
\\
&&Z=\int D\xi \exp(-S_{\rm eff}[\xi,h]). 
\label{SeffZ}\eea
We see that the metric $h$ is playing the role of the external 
scalar field in the flat space $(\tilde\tau,\vec x)$.
The calculations of the $Z$ is reduced to 
finding a saddle-point $\xi_0(h)$ from the equation 
\bea
\frac{\delta S_{\rm eff}[\xi,h]}{\delta\xi}=0 .
\label{saddle}\eea 
The explicit form of the saddle-point equation~\re{saddle} is given by
\bea
-\frac{1}{m^2}\left(\frac{d^2\xi}{d\tilde\tau^2} 
+\frac{1}{4}\left(\frac{d\xi}{d\tilde\tau}\right)^2\right)\exp(-\frac{\xi}{2})
  + (\xi -2h)=0,
\label{saddle1}\eea
 where $h=h(\tilde\tau)$ and $m$ is the scalar glueball mass. The ILM estimation 
 of the mass $m$~\cite{Tichy:2009lgv} gave
\bea
m =2.16/\rho,
\label{m}\eea 
which at $\rho\approx 0.30$~fm gives $m\approx 1290$~MeV. This one is 
in good consistency with lattice results~\cite{Meyer:2004jc} $m\approx 1470$~MeV.
 
It is natural to assume that  $H=\frac{1}{2}\frac{dh}{dt}$ 
is much less than scalar glueball mass $m$ ($H/m\ll 1$). Then, we may find
 the solution of the Eq.~\re{saddle1} as a series on $\frac{H^2}{m^2} $.
With the account of the zero and next order we have $\xi(h)=\xi_0(h)+\xi_1(h)$ where
\bea
\xi_0(h)=2h,\,\,\,\xi_1(h)=\frac{1}{m^2}
\left[2\frac{d^2 h}{d\tilde\tau^2} 
+ \left(\frac{dh}{d\tilde\tau}\right)^2\right]\exp(-h)=
\frac{1}{m^2}
\left[2\frac{d^2 h}{d\tilde t^2} 
+ \left(\frac{dh}{d\tilde t}\right)^2\right]
=-\frac{1}{m^2}\left(\frac{d H(t)}{dt} + \frac{H^2(t)}{4}\right) .
\label{saddle2}\eea
 {
The standard calculations of energy-momentum tensor  $T_{\mu\nu}$ 
from the action 
\bea
S=\int d^3xdt L(\phi(t),\dot\phi(t)),\,\,\,
L=\frac{1}{2}{\dot{\phi(t)}}^2 -V(\phi)
\eea
give 
the nontrivial components of energy-momentum  tensor as
 \bea
T_{00}= \frac{1}{2}{\dot{\phi(t)}}^2 +V(\phi),\,\,\,
T_{ij}=[\frac{1}{2}{\dot{\phi(t)}}^2 -V(\phi)]\delta_{ij}
\label{T}\eea
where $\phi(t)$ must be taken from the solution of the equations of motion 
$\frac{\delta S}{\delta\phi}=0 .$
Now we have to calculate the YM fields vacuum energy density 
$\epsilon_{\rm vac}(h)$ and
pressure $p_{\rm vac}(h)$ of the FLRW universe from the Eqs.~\re{SeffZ}, \re{saddle2},   
where we have to turn to the Minkowski signature 
and the time $dt=\exp(h/2) d\tau$ as
\bea\label{SefM}
&&S_{M,eff}[\xi,h]=\int d^3x d\tau  L_{M,eff}(\xi,h)
=\int d^3x dt\exp(-h/2)  L_{M,eff}(\xi,h),
\\
&&L_{M,eff}(\xi,h)=\frac{|\sigma_0|}{4} \frac{1}{2m^2}
 (\frac{d\xi}{dt})^2\exp(h+\frac{\xi}{2}) -V_{\rm eff}(\xi,h),\,\,\,
V_{\rm eff}(\xi,h)=\frac{|\sigma_0|}{4}(\xi-1 - 2h)\exp\xi .
\label{Lxi1}\eea 
Comparing with the standard calculations~\re{T} and using 
the solution of the saddle-point Eq.~\re{saddle2}
 we have YM fields 
vacuum energy density and pressure of FLRW universe as 
 \bea
T_{00}(h) =\epsilon_{\rm vac}(h)
&=& \exp(-h/2)\left[\frac{|\sigma_0|}{4m^2} \frac{1}{2} 
 \left(\frac{d\xi(h)}{dt}\right)^2 \exp(h+\frac{\xi(h)}{2})
 +V_{\rm eff}(\xi(h),h)\right]
 \\\nonumber
 &=&\frac{|\sigma_0|}{4} \left( \frac{H^2(t)}{2m^2} 
 -1 \right) \exp(3h(t)/2) +O\left(\frac{H^4}{m^4}\right)   ,
 \\
T_{ij}(h)=\delta_{ij}p_{\rm vac}(h),\,\,
 p_{\rm vac}(h)&=& \exp(-h/2)\left[\frac{|\sigma_0|}{4m^2} \frac{1}{2} 
 \left(\frac{d\xi(h)}{dt}\right)^2 \exp(h+\frac{\xi(h)}{2})
 -V_{\rm eff}(\xi(h),h)\right]
 \\\nonumber
 &=&\frac{|\sigma_0|}{4} \left( \frac{H^2(t)}{2m^2} 
 +1   \right)\exp(3h(t)/2)  +O\left(\frac{H^4}{m^4}\right)  
\eea
since
$T^{\rm dark}_{\mu\nu}(h)=T_{\mu\nu}(h)-T_{\mu\nu}(0)$ we have the gluon 
contributions to the dark energy density -- pressure as
\bea\label{dark}
\epsilon_{\rm dark}(h(t)) =\frac{|\sigma_0|}{4} 
\left(  \frac{H^2(t)}{2m^2}
 -1\right)(\exp(3h(t)/2)-1)+O\left(\frac{H^4}{m^4}\right)   ,
 \\\nonumber
 p_{\rm dark}(h(t)) =\frac{|\sigma_0|}{4} 
\left(  \frac{H^2(t)}{2m^2}
 +1\right)(\exp(3h(t)/2)-1)+O\left(\frac{H^4}{m^4}\right)    , 
\eea
and the equation-of-state 
\bea
w(t)=p_{\rm dark}(h(t))/\epsilon_{\rm dark}(h(t))=
-\frac{1+ \frac{H^2(t)}{2m^2} }{1- \frac{H^2(t)}{2m^2}}\le-1.
\label{w}\eea
We see from 
the Eq.~\re{dark} that at the present time $t_0$, where $h(t_0)$ is small and negative and $\frac{H^2(t_0)}{m^2} << 1$,
 the main contribution to the $\epsilon_{\rm dark}(h(t_0))$ and $p_{\rm dark}(h(t_0))$
is given by the potential term $(V_{\rm eff}(\xi(h),h) - V_{\rm eff}(0,0))$ as
\bea\label{dark1}
\epsilon_{\rm dark}(h(t_0)) = 
\frac{|\sigma_0|}{4} (1-\exp(3h(t_0)/2))+ O\left(\frac{H^2(t_0)}{m^2}\right) ,
 \\\nonumber
 p_{\rm dark}(h(t_0)) =
 -\frac{|\sigma_0|}{4} (1-\exp(3h(t_0)/2))+ O\left(\frac{H^2(t_0)}{m^2}\right)  . 
\eea
Also, accordingly Eqs.~(\ref{w}, \ref{dark1})
the present time equation-of-state  is given  
by 
\bea
w(t_0)=p_{\rm dark}(h(t_0))/\epsilon_{\rm dark}(h(t_0))=-1.
\label{w0}
\eea
The data (in MeV) are
\bea
\nonumber
&& 
\frac{|\sigma_0|}{4}=|\epsilon_{\rm vac}(0)|= 550\,{\rm MeV}/fm^3
=4.4\times 10^9\,{\rm MeV}^4,\,\,\,m=1200\,{\rm MeV},\,\,\, 
\\
&&H(t_0)=1.4\times 10^{-39}\,{\rm MeV},\,\,\, G=M_{\rm Pl}^{-2},\,\,\,
M_{\rm Pl} = 1.2 \times 10^{22}~{\rm MeV}.
\label{data} \eea 
Here $|\epsilon_{\rm vac}|,\,\,m$ are given by ILM, 
while $H(t_0),\,M_{\rm Pl}$ are present astrophysical 
data~\cite{ParticleDataGroup:2024cfk}.

The estimations (in MeV) give the present status of the FLRW metric and of the equation-of-state parameter. 

Friedmann equation~\re{Friedmann} for the metric $h(t)$ with the account of 
the Eq.~\re{dark} looks like 
\bea
\dot h^2(t)
= \frac{8\pi G|\sigma_0|}{3} (1-\exp(3h(t)/2))
+O\left(\frac{\dot h^4}{4m^4}\right) ,
\label{Friedmann1}\eea
where we neglected by the factor $\frac{\pi G|\sigma_0|}{3m^2}<<1$.
This Eq. is requesting $h(t)\le 0$ and in useful notation looks like
\bea\label{Friedmann2}
\dot f(t)=\alpha (1-\exp(f(t)))^{1/2},\,\,\,
f(t)=3h(t)/2,\,\,\,
 \alpha=3/2 \left(\frac{8\pi G|\sigma_0|}{3}\right)^{1/2}
\eea
The solution of this Eq. is given by
\bea
 (1-\exp f(t))^{1/2}=\tanh(-\alpha t/2-C)
\eea
At present time $t_0$ the metric is close to Minkowski one, which means  
$f(t_0)\sim h(t_0)\approx 0$ and
$H(t_0)\sim\alpha\tanh(-\alpha t_0/2-C)\approx 0$, $(-\alpha t_0/2-C)\approx 0$.
\\
At the initial time $ f(t_{\rm in})\sim h(t_{\rm in})\rightarrow -\infty$  means
$\tanh(-\alpha (t_{\rm in}-t_0)/2)\approx 1$. This one  corresponds to 
$\alpha (t_{\rm in}-t_0)\rightarrow -\infty.$
Finally the solution is given by
\bea\label{Friedmann3}
\frac{3}{2}\dot h(t)=\alpha (1-\exp(3h(t)/2))^{1/2}
=\alpha \tanh(-\alpha (t-t_0)/2)
\eea
The solution~\re{Friedmann3} of Eqs.~\re{Friedmann1} and \re{Friedmann2}
 at the present time $t_0$ 
with the value of $H(t_0)$ from~\re{data}
gives the present time metric
\bea\label{h0}
h(t_0)=-\frac{H^2(t_0)}{\pi G|\sigma_0|}
=-5\times 10^{-45}
\eea
At the initial time $t_{\rm in}$, when
the spatial size of the FLRW universe 
$a(t_{\rm in})\rightarrow 0\sim h(t_{\rm in})\rightarrow-\infty$, 
we have from Eqs.~\re{dark}, \re{Friedmann1}, \re{Friedmann2}, and~\re{Friedmann3}
\bea\label{Hin}
H(t_{\rm in})
=
 \left(\frac{2\pi G|\sigma_0|}{3}\right)^{1/2}
 = 1.6\times 10^{-17}\,{\rm MeV} \sim 10^{22}H(t_{0}).\,\,\,\,
\eea  
The  lifetime scale of the universe is given by
\bea\label{scale}
2/\alpha
=0.42 \times 10^{17}\,{\rm MeV^{-1}}\approx 3\times 10^{10}\,{\rm sec}
\approx 10^3\, {\rm years},
\eea 
which is too small. 
We also see that $w(t\le t_0)=-1$ with high accuracy. 
}
\section{Conclusion}

The above given QCD ILM calculations~\re{w} lead to the 
equation-of-state parameters  
$w_0=-1,\,\,\,\,w_a=0$ in accordance with $\lambda$CDM.
 {
The detailed study of the solution of Friedmann 
equations~\re{Friedmann1} and \re{Friedmann2} demonstrated 
that the contribution to the dark energy from the QCD vacuum excitations, 
generated by the universe metric, hardly plays a central role
in the dynamics of the universe, since their timescale~\re{scale} is too small.   
}
As mentioned in Sec.\ref{intro} the newest data, 
analyzed at~\cite{Shajib:2025tpd},  
prefer a dark energy from ultralight scalar field, 
corresponding to a current value
of the equation-of-state parameter $w_0\approx-0.9$~~\cite{Shajib:2025tpd}.
Such a small scalar mass is a natural in models of QCD axions.
 {
Another very interesting alternative~\cite{VanWaerbeke:2025shm} 
of the understanding of the current data 
is based on the topological properties of the QCD in a 
curved space~\cite{Urban:2009vy,Urban:2009yg,Zhitnitsky:2013pna,Zhitnitsky:2015dia,Barvinsky:2017lfl,VanWaerbeke:2025shm}, where it was shown that
YM topological configurations with the nontrivial holonomy, 
compactified to a Euclidean time circle $\mathbb{S}^1_{\kappa^{-1}}$
with the radius $\kappa^{-1}$  is providing
$O(\kappa\Lambda^3_{\rm QCD})$ contribution to the dark energy density. 
}

\section*{Acknowledgments}

I am very grateful to the referee of my work 
for the valuable suggestions and critics, 
which  improve the 
quality of my work very much. 
Also, I thank B.~Faizullaev for many fruitful discussions.
 
The work was partially supported by the 
Ministry of Higher Education, Science, and
Innovation of the Republic of Uzbekistan 
(Grant No. ALM-2023-1006-2528).

\end{document}